\documentclass[aps,pra,twocolumn,color,superscriptaddress,floatfix,nofootinbib]{revtex4-2}
\usepackage[english]{babel} %%% 'french', 'german', 'spanish', 'danish', etc.
\usepackage{amssymb}
\usepackage{amsmath}
\usepackage{mathtools}
\usepackage{txfonts}
\usepackage{mathdots}
\usepackage[normalem]{ulem}
\usepackage[dvips]{graphicx}
\usepackage{epsfig}
\usepackage{graphicx}
\usepackage{array}
\usepackage{amssymb}
\usepackage{amsfonts}
\usepackage{amsmath}
\usepackage{mathrsfs}
\usepackage{booktabs}
\usepackage{threeparttable}
\usepackage{multirow}
\usepackage{subfigure}
\usepackage{epsfig}
\usepackage{threeparttable}
\usepackage{chngpage}
\usepackage{float}
\usepackage{xcolor}
\usepackage{bm}
\usepackage{graphicx}
\usepackage[toc]{appendix}
\usepackage{tabularx}
\usepackage{graphicx}
\usepackage{adjustbox}
\usepackage{comment}
\usepackage{hyperref}
\usepackage{tabularx}
\usepackage{graphicx}
\usepackage{adjustbox}

\hypersetup{
	unicode=false,     % non-Latin characters in Acrobat bookmarks
	pdftoolbar=false,  % show Acrobat toolbar?
	pdfmenubar=true,   % show Acrobat menu?
	pdffitwindow=false, % window fit to page when opened
	pdfstartview={FitH},% fits the width of the page to the window
	pdftitle={},    % title
	pdfauthor={Authors},     % author
	pdfsubject={},   % subject of the document
	pdfcreator={},   % creator of the document
	pdfproducer={}, % producer of the document
	pdfkeywords={} {} {}, % list of keywords
	pdfnewwindow=true,% links in new window
	colorlinks=true,% false: boxed links; true: colored links
	linkcolor=black,% color of internal links (change box color with linkbordercolor)
	citecolor=blue, % color of links to bibliography
	filecolor=magenta,% color of file links
	urlcolor=blue% color of external links
}

\makeatletter

\newcommand{\Rmnum}[1]{\expandafter\@slowromancap\romannumeral #1@}
\makeatother

\begin{document}
	
	\title{Temperature effect on a kicked Tonks-Girardeau gas}
	
	\author{Ang Yang} 
	\thanks{These authors contributed equally.}
	\affiliation{School of Physics, Zhejiang Key Laboratory of Micro-nano Quantum Chips and Quantum Control, Zhejiang University, Hangzhou $310027$, China}
	
	\author{Yue Chen} 
	\thanks{These authors contributed equally.}
	\affiliation{School of Physics, Zhejiang Key Laboratory of Micro-nano Quantum Chips and Quantum Control, Zhejiang University, Hangzhou $310027$, China}
	
	\author{Lei Ying} \email{leiying@zju.edu.cn}
	\affiliation{School of Physics, Zhejiang Key Laboratory of Micro-nano Quantum Chips and Quantum Control, Zhejiang University, Hangzhou $310027$, China}
	
	\begin{abstract}
		It is widely recognized that finite temperatures degrade quantum coherence and can induce thermalization. Here, we study the effect of finite temperature on a kicked Tonks-Girardeau gas, which is known to exhibit many-body dynamical localization and delocalization under periodic and quasiperiodic kicks, respectively. We find that many-body dynamical localization persists at finite—and even high—temperatures, although the coherence of the localized state is further degraded. In particular, we demonstrate a modified effective thermalization of the localized state by considering the initial temperature. Moreover, we show many-body dynamical localization transition at intermediate temperature. Our work extends the study of many-body dynamical localization and delocalization to the finite-temperature regime, providing guidance for cold-atom experiments, particularly in the strongly-interacting regime.
	\end{abstract}
	
	\date{\today}
	\maketitle
	
	\section{Introduction}
	Conventional wisdom suggests that continuously injecting energy into a system inevitably leads to thermalization and the loss of its initial-state memory. However, quantum coherence brings new possibilities. A textbook counterexample is the quantum kicked rotor (QKR), in which the kinetic energy of a free particle, instead of growing under continuous periodic kicks, gradually saturates~\cite{casati1979quantumpendulum,fishman1982quantumrecurrence,grempel1984nonintegrable,chirikov1986localizationofchaos,Altland1993qkr,Altland1996qftqkr,SANTHANAM2022rotorreview}. This phenomenon occurs in momentum space and is called dynamical localization (DL), analogous to the well-known Anderson localization (AL) in real space~\cite{anderson1958absenceofdiffusion}. A key signature of the DL is the nearly exponential localization of the wave function in momentum space and this phenomenon has been observed and extensively studied over recent decades~\cite{moore1994DL,moore1995rotorexp,raizen1998noiseDL,christensen1998decoherence,darcy2001diffusionexp}. Moreover, many variants of QKR are introduced to explore new phenomena, such as the quasiperiodic QKR~\cite{casatiquasi1989,delande2008metalinsulatorexp,delande2009metalinsulator,tian_quasi2011}, the spin-$1/2$ QKR~\cite{chen_spin_qkr_2014,tian_spin_qkr2016}, the double kicked rotor~\cite{wang_double_qkr_2008,zhou_double_qkr_2018}, the parity-time symmetric QKR~\cite{prosen_PT_qkr_2010,stefano_PT_qkr_2017,li_PT_qkr_2024}. 
	
	Furthermore, various many-body versions of the QKR with different interactions have been proposed to investigate the persistence of DL~\cite{shepelyansky1993nonlinear,shepelyansky2008nonlinearity,Gligoric2011delocalization,qin2017interacting,victor2017couplerelativistic,victor2016integrableDMBL,fazio2018CQKRdiffusion,konik2020MBDL,delande2020meanfield,vuatelet2021effectivethermal,chicireanu2021fewbodylimit,fazio2021CQKRsubdiffusion,Vuatelet2023dynamicalmanybody,olsen2025anderson,yang2025origin}. Among them, the kicked Lieb-Liniger (LL) model has attracted widespread attention due to its high experimental feasibility and is predicted to exhibit many-body dynamical localization (MBDL) as well as delocalization. Recently, a series of cold-atom experiments based on the kicked LL model have verified these phenomena~\cite{david2022delocalization,gupta2022delocalization,guo2023observation,gupta2024anderson_transition}.
	
    However, finite-temperature effects in the QKR model have not yet been studied, as most of existing experimental and theoretical works are limited to near-zero temperature settings. In this paper, we investigate the role of finite temperatures by focusing on the  LL model in the limit of infinite interactions and finite temperatures, under both periodic and quasiperiodic kicks, i.e., the kicked thermal Tonks-Girardeau (TG) gas~\cite{girardeau1960_TGgas,schultz1963_TGgas,lenard1964_TGgas,lenard1966_thermalTG,paredes2004tonks,girardeau2005FBmapping,buljan2008bosefermimap}. For the periodic case, we find MBDL always persists with higher localized energy and the coherence of localized state is further reduced, as the initial temperature increases. We further demonstrate the effective thermalization of the MBDL state at low and high temperatures by establishing a nonlinear relationship between the effective temperature and the localized length. For the quasiperiodic case, a many-body dynamical localization-delocalization transition is discovered at an intermediate temperature. We quantify different phases by one-parameter scaling laws. In particular, the effective thermalization of the delocalized state is studied.
	
	The paper is organized as follows. The model and prerequisites are first introduced in Sec. II. In Sec. III, we introduce the grand-ensemble lattice approach and some necessary observables like the one-particle density matrix (OPDM). The periodically and quasiperiodically kicked TG gas are studied in Sec. IV and V, respectively.

	\section{Model}
	We consider the kicked Lieb-Liniger (LL) model~\cite{konik2020MBDL,chicireanu2021fewbodylimit,vuatelet2021effectivethermal,Vuatelet2023dynamicalmanybody}, described as $N$ bosons of mass $M$ with contact interactions moving in a one-dimensional space of length $L$, subjected to a pulsed cosine potential. The dimensionless model Hamiltonian consists of two parts:
	\begin{equation}
		\hat{H}(t) = \hat{H}_{\mathrm{LL}} + \sum_{n\in\mathbb{N}}\delta(t-n)\hat{H}_{\mathrm{K}},
		\label{total_hamiltonian}
	\end{equation}
	where the LL Hamiltonian is~\cite{lieb1963solution1}
	\begin{equation}
		\hat{H}_{\mathrm{LL}} = \sum_i^N\frac{\hat{p}_i^2}{2} + g\sum^N_{i<j}\delta\left(\hat{x}_i-\hat{x}_j\right) ,
		\label{LL_hamiltonian}
	\end{equation}
	and the kick Hamiltonian is
	\begin{equation}
		\hat{H}_{\mathrm{K}} = \mathcal{K}(t)\sum_i^N \mathrm{cos}\left(\hat{x}_i\right),
		\label{kick_hamiltonian}
	\end{equation}
	with $\mathcal{K}(t) = K\left[1+\varepsilon\cos{(\omega_2t)}\cos{(\omega_3t)}\right]$. All units are rescaled to be dimensionless such that time is in units of the pulse period $\tau$ and length in units of the inverse of the kick-potential wave number $k_{\mathrm{L}}$~\cite{delande2009metalinsulator,vuatelet2021effectivethermal}. The commutation relation now reads $[\hat{x}_i,\hat{p}_j]=i\hbar_{\rm{eff}}\delta_{ij}$, where the effective reduced Planck constant is $\hbar_{\rm{eff}} = 4\tau\hbar k_{\mathrm{L}}^2/M$. The characteristic parameters $g, K$ and $\varepsilon$ denote the interaction strength, kick strength, and anisotropy strength, respectively. The incommensurate frequencies are fixed respectively as $\omega_2=2\pi\sqrt{5}$ and $ \omega_3=2\pi\sqrt{13}$. In the following text, we  set $L=2\pi$, and the Boltzmann constant is fixed as $k_{\mathrm{B}}=1$. Unless stated otherwise, all our results presented in the following sections are obtained in the Tonks-Girardeau (TG) limit $g\rightarrow\infty$.
	
	We firstly recall the zero-temperature $(T=0)$ results at two limits $g=0$ and $g \to \infty$. For $g=0$ and $\varepsilon=0$, we recover Quantum kicked rotor (QKR) displaying dynamical localization~\cite{casati1979quantumpendulum,fishman1982quantumrecurrence,grempel1984nonintegrable}, while for $g=0$ and $\varepsilon\neq0$, our model reduces to the quasi-periodic QKR with an Anderson-like transition at large $K, \varepsilon$~\cite{casatiquasi1989,delande2009metalinsulator}. At the opposite limit $g \to \infty$, the bosons are impenetrable and form the TG gas~\cite{girardeau1960_TGgas}. In this regime, according to the Bose-Fermi correspondence, one can explicitly obtain the many-body bosonic wavefunction as
	\begin{equation}
		\Psi_{\mathrm{B}}({x},t)=\prod_{i>j}(x_i-x_j)\Psi_{\mathrm{F}}({x},t),
	\end{equation}
	where $\Psi_{\mathrm{F}}({x},t)=(L^N N!)^{-1/2}\mathrm{det}\big[\psi_i(x_j,t)\big]$ is the many-body fermionic wavefunction in terms of Slater's determinant constructed by the single-particle orbitals $\psi_i(x_j,t)$. Thus, all the local observables (such as energy, entropy, density, etc.) of Tonks gas are the same as those of free fermions, except for the nonlocal quantities like one-particle density matrix (OPDM)~\cite{girardeau1960_TGgas,lenard1964_TGgas}. For a kicked Tonks gas without anisotropy, many-body dynamical localization persists, accompanied by an algebraic tail $k^{-4}$ and effective thermalization~\cite{konik2020MBDL,chicireanu2021fewbodylimit,vuatelet2021effectivethermal}. For a kicked Tonks gas with anisotropy, a many-body dynamical localization-delocalization transition exists with the same phase diagram as for free fermions~\cite{Vuatelet2023dynamicalmanybody}.
	
	Intuitively, one might expect finite temperature to primarily affect inter-boson coherence, as the OPDM of a TG gas decays exponentially for $T>0$ but algebraically at $T=0$~\cite{lenard1964_TGgas,lenard1966_thermalTG,tracy1979longrange,rigol2005groundstate,rigol2005_thermalTG}. It was shown that the kicks also destroy the coherence of quasicondensate~\cite{vuatelet2021effectivethermal}. We therefore expect dynamical localization to persist at finite temperatures.
	
	\section{Numerical approach}
    %%%%%%%%%%%%%%%%%%%%%%%%%%%%%%%%%%%%%%%%%%%%%%%%%%%%%%%%%%%%%
	\begin{figure*}[ht]
		\centering
		\includegraphics[width=0.7\linewidth]{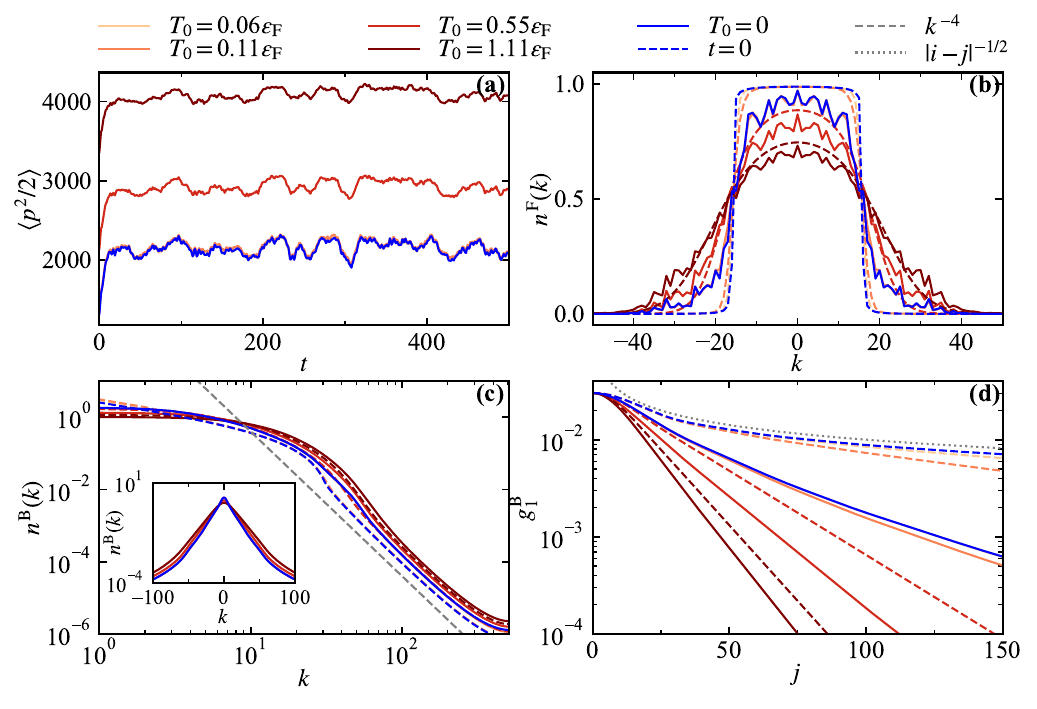}
		\caption{
			(a) The kinetic energy dynamics of the periodically kicked Tonks gas for different initial temperature $T_0$. (b-d) The corresponding fermionic momentum distribution (b), the bosonic momentum distribution (c), and the bosonic correlation function (d) for different initial temperature $T_0$. The dashed lines (blue and red) denote the initial state at $t=0$ while the solid lines denote the final state at $t=500$. The particle number, anisotropy, kick strength and effective Planck constant are respectively $N=31, \varepsilon=0, K=4, \hbar_{\mathrm{eff}}=1$.
		}
		\label{periodic_dynamics}
	\end{figure*}
	%%%%%%%%%%%%%%%%%%%%%%%%%%%%%%%%%%%%%%%%%%%%%%%%%%%%%%%%%%%%%
	Finite temperatures can be naturally incorporated using the grand-canonical ensemble. Here we follow the exact lattice method of Refs.~\cite{rigol2005_thermalTG,rigol2017thermalexpansion} to treat the thermal Tonks-Girardeau (TG) gas. At the TG limit $g\rightarrow\infty$, after the spatial discretization,
	Eqs.~\eqref{LL_hamiltonian} and \eqref{kick_hamiltonian} are rewritten as the hard-core Bose-Hubbard Hamiltonian
	\begin{equation}
		\begin{aligned}
			\hat{H}_{\mathrm{LL}}&=-J\sum_{l=1}^{\mathcal{N}-1}\left(\hat{b}_{l+1}^{\dagger}\hat{b}_l+\mathrm{H.c.}\right)+2J\sum_{l=1}^{\mathcal{N}}\hat{n}_l ,\\
			\hat{H}_{\mathrm{K}}&=\mathcal{K}(t)\sum_{l=1}^{\mathcal{N}}\mathrm{cos}\left(la-\frac{\mathcal{N}+1}{2}a\right)\hat{n}_l,
			\label{hcb_hamiltonian}
		\end{aligned}
	\end{equation}
	where $\mathcal{N}$ (odd), $a=\frac{L}{\mathcal{N}-1}$ and $J=\frac{\hbar^2_{\mathrm{eff}}}{2a^2}$ are the number of discrete lattice sites, the lattice spacing, and the effective coupling constant, respectively. The bosonic creation $\hat{b}_l^{\dag}$ and annihilation $\hat{b}_l$ operators satisfy the commutation relation $[\hat{b}_l,\hat{b}_j^{\dagger}]=\delta_{lj}$ with the additional on-site constraints $(\hat{b}_l^{\dagger})^2=\hat{b}_l^2=0$. The local density operator is $\hat{n}_l=\hat{b}_l^{\dagger}\hat{b}_l$ and particle-number conservation requires $\sum_{l=1}^{\mathcal{N}} \langle \hat{n}_l\rangle=N$. Here, we consider the low fillings case ($N/\mathcal{N}\rightarrow0$), so the system is effectively in the continuum. For computational convenience, we set $\mathcal{N}=2^{10}$ and $N<101$ throughout this paper.
	
	To study the dynamics of a kicked thermal Tonks gas, we compute the equal-time one-particle density matrix (OPDM) of bosons at time $t$ as
	\begin{equation}
		\rho^{\mathrm{B}}_{ij}(t)=\langle \hat{b}_i^{\dagger}(t)\hat{b}_j(t)\rangle=\mathrm{Tr}\left[\hat{U}(t)\hat{\rho}_\mathrm{LL}\hat{U}^{\dagger}(t)\hat{b}_i^{\dagger}\hat{b}_j\right],
		\label{opdm}
	\end{equation}
	where $\hat{\rho}_{\mathrm{LL}}=e^{-(\hat{H}_{\mathrm{LL}}-\mu N)/T}/Z$ is the initial thermal density matrix at temperature $T$ with the chemical potential $\mu$, the partition function $Z=\mathrm{Tr}[e^{-(\hat{H}_{\mathrm{LL}}-\mu N)/T}]$ and the unitary evolution operator is $\hat{U}(t)=(e^{-i\hat{H}_{\mathrm{LL}}/\hbar_\mathrm{eff}}e^{-i\hat{H}_{\mathrm{K}}/\hbar_\mathrm{eff}})^t$. To evaluate Eq.~\eqref{opdm}, we map the hard-core bosons onto non-interacting spinless fermions by the Jordan-Wigner transformation~\cite{jordan1928} 
	\begin{equation}
		\hat{b}_l^{\dagger}=\hat{f}_l^{\dagger}\prod_{\beta=1}^{l-1}e^{-i\pi\hat{f}_{\beta}^{\dagger}\hat{f}_{\beta}},\quad 
		\hat{b}_l=\prod_{\beta=1}^{l-1}e^{i\pi\hat{f}_{\beta}^{\dagger}\hat{f}_{\beta}}\hat{f}_l.
	\end{equation}
	The corresponding fermionic Hamiltonian form is the same as Eqs.~\eqref{hcb_hamiltonian} by replacing $\hat{b}_l^{\dagger}$ $(\hat{b}_l)$ with $\hat{f}_l^{\dagger}$ $(\hat{f}_l)$. This transformation allows us to leverage identities in the fermionic Fock space to simplify the expression for the OPDM as~\cite{rigol2005_thermalTG}
	\begin{equation}
		\begin{aligned}
			\rho^{\mathrm{B}}_{ij}(t)=&\frac{1}{Z}\left\{\mathrm{det}[I+\hat{U}(t)\hat{\rho}_\mathrm{LL}\hat{U}^{\dagger}(t)\hat{O}_i\hat{O}_j+\hat{U}(t)\hat{\rho}_\mathrm{LL}\hat{U}^{\dagger}(t)\hat{O}_i\hat{A}\hat{O}_j]\right. \\
			&\left.-\mathrm{det}[I+\hat{U}(t)\hat{\rho}_\mathrm{LL}\hat{U}^{\dagger}(t)\hat{O}_i\hat{O}_j]\right\},\quad \mathrm{for}\; i\neq j \\
			\rho^{\mathrm{B}}_{ii}(t)=&1-{\left[I+e^{-(\hat{H}_{\mathrm{LL}}-\mu I)/T}\right]}^{-1}_{ii},
		\end{aligned}
	\end{equation}
	where all the involved Hamiltonians are reduced to one-particle case with a dimension $\mathcal{N\times\mathcal{N}}$, matrix $\hat{A}$ has only one non-zero element $A_{ij}=1$ and matrix $\hat{O}_i$ $(\hat{O}_j)$ is diagonal with the first $i-1$ $(j-1)$ elements equal to $-1$ and the rest equal to $1$. The chemical potential $\mu$ is determined by the constraint
	\begin{equation}
		\sum_{i=1}^{\mathcal{N}}\rho^{\mathrm{B}}_{ii}=\sum_{i=1}^{\mathcal{N}}\frac{1}{e^{(E^i_{\mathrm{LL}}-\mu)/T}+1}=N,
	\end{equation}
	which is the Fermi-Dirac distribution and $E^i_{\mathrm{LL}}$ is the eigenvalue of $\hat{H}_{\mathrm{LL}}$. Note that, for periodic boundary conditions on the bosons, the mapped fermions obey periodic boundary conditions when $N$ is odd and anti-periodic boundary conditions when $N$ is even~\cite{girardeau1960_TGgas,lieb1963solution1}. Since the particle-number fluctuates around an average value in the grand-canonical ensemble, only the open boundary condition is considered here. Using OPDM, the bosonic momentum distribution can be obtained as
	\begin{equation}
		n^{\mathrm{B}}(k,t)=\frac{1}{\mathcal{N}-1}\sum_{j,l=1}^{\mathcal{N}}e^{-ik(j-l)a}\rho^{\mathrm{B}}_{jl}(t),
	\end{equation}
	and the correlation function is given by
	\begin{equation}
		g^{\mathrm{B}}_1(j,t) = \frac{1}{\mathcal{N}}\sum_{i=1}^{\mathcal{N}} \rho^{\mathrm{B}}_{i,i+j}(t).
	\end{equation}
	Similarly, we can obtain the fermionic one-particle density matrix $\rho^{\mathrm{F}}_{ij}(t)$ and the fermionic momentum distribution $n^{\mathrm{F}}(k,t)$.
	
	\section{Periodic case}
	\subsection{MBDL at finite temperature}
	In this section, we consider the periodically kicked Tonks gas and set $\varepsilon=0, \hbar_{\mathrm{eff}}=1$. We distinguish between high and low temperature regions in terms of the Fermi energy $\varepsilon_\mathrm{F}$ with $\varepsilon_\mathrm{F}=\hbar_{\mathrm{eff}}^2N^2/8$ in our units. Starting from the ground state of the Tonks gas at five different initial temperatures $T_0$, the kinetic energy growth of the Tonks gas after $500$ kicks is shown in Fig.~\ref{periodic_dynamics}(a). Clearly, all the kinetic energy saturates after tens of kicks and consistently fluctuates around a certain value hereafter, indicating the persistence of the many-body dynamical localization (MBDL) at finite and even high temperature ($T_0\sim\varepsilon_\mathrm{F}$). At a very low initial temperature ($T_0=0.06\varepsilon_\mathrm{F}$), the energy growth almost coincides with that at zero temperature. As $T_0$ increases, the kinetic energy after saturation is higher, while the overall growth trend remains similar. At higher $T_0$, the small kinetic energy fluctuations in the localized regime are smoothed out, indicating that finite temperatures further degrade the coherence of the thermal Tonks-Girardeau (TG) gas. In the Tonks regime, the bosons and underlying fermions share the same eigen-spectrum, thus the same energy dynamics. As shown in Fig.~\ref{periodic_dynamics}(b), all the fermionic momentum distributions $n^{\mathrm{F}}(k)$ initially obey the Fermi-Dirac distribution (dashed lines) and they are expanded after hundreds of kicks (solid lines). For the same kick strength $K$, $n^{\mathrm{F}}(k)$ at low $T_0$ is more broadened than at high $T_0$. This is because the kicks transfer the fermions from the low-momentum states to higher ones, which are already occupied at finite temperature. Consequently, finite temperatures appear to enhance the robustness of the TG gas against external perturbations by limiting available phase space.
	
	One of the most important signatures of the MBDL is that the steady-state bosonic momentum distribution decays exponentially ($\propto e^{-k/k_{\mathrm{loc}}}$) at low momenta but algebraically ($\propto \mathcal{C}_{\mathrm{ss}}/k^{4}$) at high momenta~\cite{chicireanu2021fewbodylimit}, as shown in Fig.~\ref{periodic_dynamics}(c). Finite temperature leads to the increase of both localization length $k_{\mathrm{loc}}$ and $\mathcal{C}_{\mathrm{ss}}$. At zero temperature, the ground state of Tonks gas is known to hold long-range coherence~\cite{lenard1964_TGgas,tracy1979longrange,rigol2005groundstate,rigol2011coldatomreview}, i.e., the correlation function decays algebraically as $g^{\mathrm{B}}_1(r)\propto r^{-1/2}$ (blue dashed line in Fig.~\ref{periodic_dynamics}(d)). The kicks destroy the long-range coherence and lead to the exponential decay of $g^{\mathrm{B}}_1(r)$ as $g^{\mathrm{B}}_1(r)\propto e^{-2r/r_{\mathrm{c}}}$ (blue solid line). At finite temperature, the initial state already lacks long-range coherence, and the kicks further suppress coherence, reducing the correlation length $r_{\mathrm{c}}$.

    %%%%%%%%%%%%%%%%%%%%%%%%%%%%%%%%%%%%%%%%%%%%%%%%%%%%%%%%%%%%%
	\begin{figure}[ht!]
		\centering
		\includegraphics[width=0.75\linewidth]{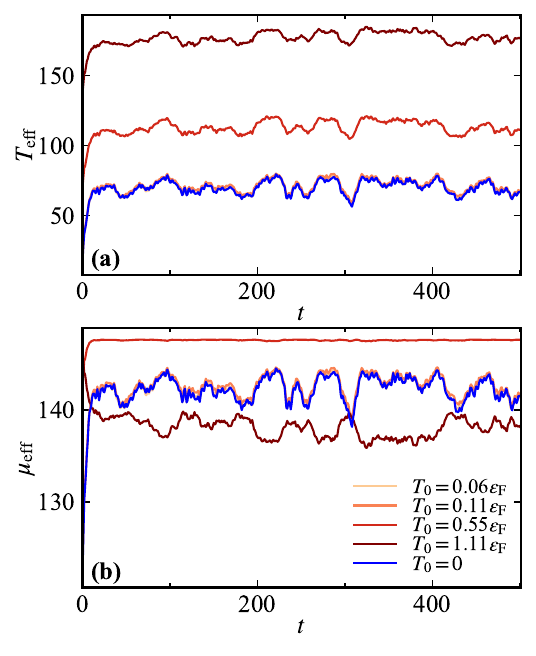}
		\caption{
			Time evolution of the extracted effective temperature $T_{\mathrm{eff}}$ (a) and effective chemical potential $\mu_{\mathrm{eff}}$ (b) for different initial temperature $T_0$. The particle number, anisotropy, kick strength and effective Planck constant are respectively $N=31, \varepsilon=0, K=4, \hbar_{\mathrm{eff}}=1$.
		}
		\label{periodic_fitting}
	\end{figure}
	%%%%%%%%%%%%%%%%%%%%%%%%%%%%%%%%%%%%%%%%%%%%%%%%%%%%%%%%%%%%%
    %%%%%%%%%%%%%%%%%%%%%%%%%%%%%%%%%%%%%%%%%%%%%%%%%%%%%%%%%%%%%
	\begin{figure}[ht!]
		\centering
		\includegraphics[width=0.75\linewidth]{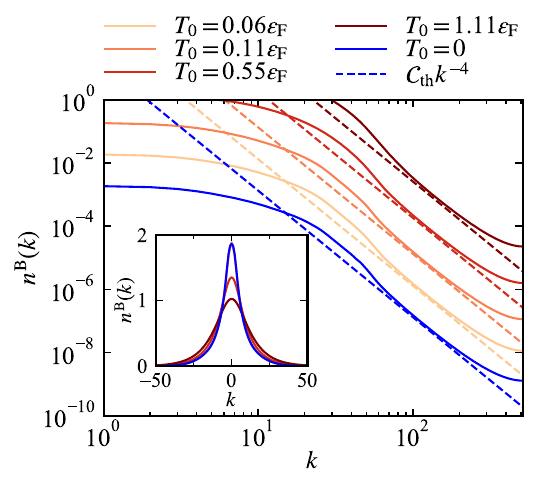}
		\caption{
			The bosonic momentum distribution at $t=500$ for different initial temperature $T_0$ in a log-log scale. The dashed lines denote the predicted algebraic decay $n^{\mathrm{B}}(k)\propto \mathcal{C}_{th}/k^{4}$ with $\mathcal{C}_{th}$ being the Tan's contact of a thermal Tonks gas. All the data have been shifted for better visibility. Inset shows the same data in a linear scale. The particle number, anisotropy, kick strength and effective Planck constant are respectively $N=31, \varepsilon=0, K=4, \hbar_{\mathrm{eff}}=1$.
		}
		\label{periodic_tan_contact}
	\end{figure}
	%%%%%%%%%%%%%%%%%%%%%%%%%%%%%%%%%%%%%%%%%%%%%%%%%%%%%%%%%%%%%
    %%%%%%%%%%%%%%%%%%%%%%%%%%%%%%%%%%%%%%%%%%%%%%%%%%%%%%%%%%%%%
	\begin{figure}
		\centering
		\includegraphics[width=0.75\linewidth]{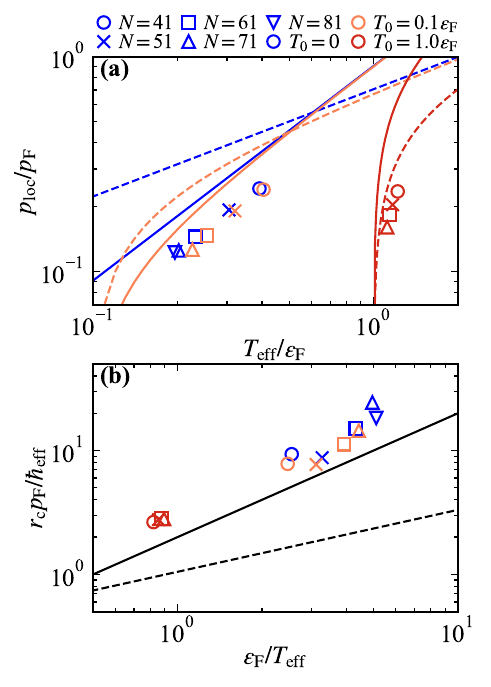}
		\caption{
			(a) $p_{\mathrm{loc}}/p_{\mathrm{F}}$ as a function of $T_{\mathrm{eff}}/\varepsilon_{\mathrm{F}}$ at $t=500$ for different particle numbers $N$ and different initial temperature $T_0$. The solid (dashed) lines denote the corresponding prediction for low (high) temperature. (b) $r_{\mathrm{c}}p_{\mathrm{F}}/\hbar_{\mathrm{eff}}$ as a function of $\varepsilon_{\mathrm{F}}/T_{\mathrm{eff}}$ at $t=500$ for different particle numbers $N$ and different initial temperature $T_0$. The black solid and dashed lines show the predictions for $\mu_{\mathrm{eff}}/T_{\mathrm{eff}}\gg 1$ and $\mu_{\mathrm{eff}}/T_{\mathrm{eff}}\ll 1$, respectively. The anisotropy, kick strength and effective Planck constant are respectively $\varepsilon=0, K=4, \hbar_{\mathrm{eff}}=1$.
		}
		\label{periodic_scaling}
	\end{figure}
	%%%%%%%%%%%%%%%%%%%%%%%%%%%%%%%%%%%%%%%%%%%%%%%%%%%%%%%%%%%%%
    \subsection{Effective thermalization of MBDL state at finite temperatures}
	It has been shown that the many-body dynamical localization (MBDL) state at zero temperature is well described by a thermal density matrix of the Tonks gas, i.e., the effective thermalization~\cite{vuatelet2021effectivethermal},
	\begin{equation}
		\hat{\rho}_{\mathrm{ss}}\simeq\hat{\rho}_{\mathrm{th}}\propto e^{-(\hat{H}_{\mathrm{TG}}-\mu_{\mathrm{eff}})/T_{\mathrm{eff}}},
	\end{equation}
	where $\mu_{\mathrm{eff}}$ and $T_{\mathrm{eff}}$ are the effective chemical potential and effective temperature, respectively. At finite temperatures, the evolved state remains qualitatively similar to a thermal Tonks-Girardeau (TG) gas at a higher temperature, as the initial coherence is already significantly reduced. We can extract $T_{\mathrm{eff}}$ and $\mu_{\mathrm{eff}}$ from the underlying fermions by fitting $n^{\mathrm{F}}(k)$ to the Fermi-Dirac distribution. The constraints are written as~\cite{vuatelet2021effectivethermal}
	\begin{equation}
		\begin{aligned}
			&\sum_k f_{\mathrm{FD}}(k, T_{\mathrm{eff}}, \mu_{\mathrm{eff}})=N, \\
			&\sum_k \frac{\hbar_{\mathrm{eff}}^2k^2}{2}f_{\mathrm{FD}}(k, T_{\mathrm{eff}}, \mu_{\mathrm{eff}})=E_{\mathrm{F}},
		\end{aligned}
		\label{constraints}
	\end{equation}
	where $E_{\mathrm{F}}=\int \frac{\hbar_{\mathrm{eff}}^2k^2}{2}n^{\mathrm{F}}(k)dk$ represents the kinetic energy of fermions and 
	\begin{equation}
		f_{\mathrm{FD}}(k, T_{\mathrm{eff}}, \mu_{\mathrm{eff}})=\frac{1}{e^{(\hbar_{\mathrm{eff}}^2k^2/2-\mu_{\mathrm{eff}})/T_{\mathrm{eff}}}+1}
	\end{equation}
	stands for the Fermi-Dirac distribution. The time evolution of the extracted $T_{\mathrm{eff}}$ and $\mu_{\mathrm{eff}}$ with kick number is shown in Fig.~\ref{periodic_fitting}. The time evolution of $T_{\mathrm{eff}}$ closely mirrors that of the kinetic energy in Fig.~\ref{periodic_dynamics}(a), confirming that effective thermalization governs the steady state. For the initial temperature $T_0$ considered here, the effective chemical potential is always positive, while $\mu_{\mathrm{eff}}$ for $T_0=1.11\varepsilon_\mathrm{F}$ drops with time. This indicates the system enters the high-temperature regime when $T_0\sim\varepsilon_\mathrm{F}$. For a thermal Tonks gas at temperature $T$, the bosonic momentum distribution is known to decay as $n^{\mathrm{B}}(k)\propto \mathcal{C}_{\mathrm{th}}k^{-4}$ with $\mathcal{C}_{\mathrm{th}}={8NE(T,N)}/{(L^2\hbar_{\mathrm{eff}}^2)}$ being the Tan's contact~\cite{vignolo2013contact,TAN20082952,TAN20082971,olshanii2003short} and $E(T,N)$ is the total energy. Thus we expect $\mathcal{C}_{\mathrm{ss}}=\mathcal{C}_{\mathrm{th}}$ in the MBDL phase. As shown in Fig.~\ref{periodic_tan_contact}, $n^{\mathrm{B}}(k)$ for different $T_0$ at high momentum can be well captured by the prediction (dashed lines).
	
	Since the effective temperature $T_{\mathrm{eff}}$ follows similar dynamics to the system's energy, we can establish a relationship between $T_{\mathrm{eff}}$ and the system's parameters. From the perspective of fermions, for low temperature ($T\ll\varepsilon_{\mathrm{F}}$), we can perform the Sommerfeld expansion to the energy of the initial and final states, respectively as
	\begin{equation}
		\begin{aligned}
			E_{\mathrm{F}}(T_0) &\simeq \frac{N\varepsilon_{\mathrm{F}}}{3}+\frac{N\pi^2}{12}\frac{T_0^2}{\varepsilon_{\mathrm{F}}}+O(T_0^4),\\
			E_{\mathrm{F}}(T_{\mathrm{eff}}) &\simeq \frac{N\varepsilon_{\mathrm{F}}}{3}+\frac{N\pi^2}{12}\frac{T_{\mathrm{eff}}^2}{\varepsilon_{\mathrm{F}}}+O(T_{\mathrm{eff}}^4).
		\end{aligned}
		\label{sommerfeld_expansion}
	\end{equation}
	In terms of the MBDL, the energy of the steady-state can also be expressed as
	\begin{equation}
		E_{\mathrm{F}}(T_{\mathrm{eff}})=E_{\mathrm{F}}(T_0)+\frac{Np_{\mathrm{loc}}^2}{2},
		\label{localized_energy}
	\end{equation}
	where $p_{\mathrm{loc}}$ is the averaged localization length of the fermions. Substituting Eq.~\eqref{sommerfeld_expansion} into Eq.~\eqref{localized_energy} and ignoring the high-order terms, we have the following scaling relation
	\begin{equation}
		\frac{p_{\mathrm{loc}}}{p_{\mathrm{F}}}=\frac{\pi}{2\sqrt{3}}\sqrt{\frac{T_{\mathrm{eff}}^2}{\varepsilon_{\mathrm{F}}^2}-\frac{T_0^2}{\varepsilon_{\mathrm{F}}^2}},
		\label{low_T_scaling}
	\end{equation}
	which is independent of the particle number $N$. Unlike the zero temperature case, if the initial $T_0$ is not negligible, the relation between $p_{\mathrm{loc}}$ and $T_{\mathrm{eff}}$ is nonlinear. In the opposite limit $(T\gg\varepsilon_{\mathrm{F}})$, assuming that the fermions are weakly degenerate, we have the energy of the initial and final states, respectively as
	\begin{equation}
		\begin{aligned}
			E_{\mathrm{F}}(T_0) &\simeq \frac{NT_0}{2}+O(T_0^{-3/2}),\\
			E_{\mathrm{F}}(T_{\mathrm{eff}}) &\simeq \frac{NT_{\mathrm{eff}}}{2}+O(T_{\mathrm{eff}}^{-3/2}),
		\end{aligned}
		\label{high_T_expansion}
	\end{equation}
	where the high-order term $O(T^{-3/2})$ is the additional energy caused by the quantum statistical correlation. Substituting Eq.~\eqref{high_T_expansion} into Eq.~\eqref{localized_energy} and ignoring the high-order terms, we expect another scaling relation
	\begin{equation}
		\frac{p_{\mathrm{loc}}}{p_{\mathrm{F}}}=\frac{1}{\sqrt{2}}\sqrt{\frac{T_{\mathrm{eff}}}{\varepsilon_{\mathrm{F}}}-\frac{T_0}{\varepsilon_{\mathrm{F}}}}.
		\label{high_T_scaling}
	\end{equation}
	In Fig.~\ref{periodic_scaling}(a), we plot $p_{\mathrm{loc}}/p_{\mathrm{F}}$ as a function of $T_{\mathrm{eff}}/\varepsilon_{\mathrm{F}}$ at the MBDL regime for different particle numbers $N$ and different initial temperature $T_0$. Evidently, all the data for different $N$ collapse to a unified function. The data at $T_0=0$ follow a linear scaling, which is well aligned with the low-temperature prediction Eq.~\eqref{low_T_scaling} (blue solid line). At very low $T_0$, the data are still captured by Eq.~\eqref{low_T_scaling} with a nonlinear increase. However, when $T_0$ is comparable to $\varepsilon_{\mathrm{F}}$, the data deviate from the low-temperature prediction and approach Eq.~\eqref{high_T_scaling} (red dashed line), as expected.
	
	From the perspective of bosons, for a thermal Tonks gas with positive chemical potential ($\mu_{\mathrm{eff}}>0$), the correlation length of the correlation function $g^{\mathrm{B}}_1$ is known to read~\cite{rigol2011coldatomreview,korepin1991correlation}
	\begin{equation}
		r_{\mathrm{c}}^{-1}=\frac{\sqrt{2k_{\mathrm{B}}T}}{2\pi\hbar_{\mathrm{eff}}}\int_{-\infty}^{\infty}d\lambda \left|\frac{e^{\lambda^2-\mu_{\mathrm{eff}}/T}+1}{e^{\lambda^2-\mu_{\mathrm{eff}}/T}-1}\right|.
	\end{equation}
	For $\mu_{\mathrm{eff}}/T_{\mathrm{eff}}\gg1$, one obtains $r_{\mathrm{c}}\sim \hbar_{\mathrm{eff}}v_{\mathrm{F}}/T_{\mathrm{eff}}$, where $v_{\mathrm{F}}$ is the Fermi velocity. Thus, we expect the following scaling in the MBDL regime
	\begin{equation}
		\frac{r_{\mathrm{c}}p_{\mathrm{F}}}{\hbar_{\mathrm{eff}}}=2\frac{\varepsilon_{\mathrm{F}}}{T_{\mathrm{eff}}}.
		\label{bosons_scaling_low_T}
	\end{equation}
	As shown in Fig.~\ref{periodic_scaling}(b), the low and zero $T_0$ cases are well captured by the above scaling (black solid line). But for $\mu_{\mathrm{eff}}/T_{\mathrm{eff}}\ll1$, we get that~\cite{korepin1991correlation}
	\begin{equation}
		\frac{r_{\mathrm{c}}p_{\mathrm{F}}}{\hbar_{\mathrm{eff}}}=\frac{\sqrt{\pi}}{(1-2^{-3/2})\zeta(3/2)}\sqrt{\frac{\varepsilon_{\mathrm{F}}}{T_{\mathrm{eff}}}},
		\label{bosons_scaling_high_T}
	\end{equation}
	where $\zeta(x)$ is the Riemann $\zeta-$function. For $T\sim1\varepsilon_{\mathrm{F}}$, we can see from Fig.~\ref{periodic_fitting}(b) that $\mu_{\mathrm{eff}}$ is smaller than but comparable to $T_{\mathrm{eff}}$, so the data partially follow Eq.~\eqref{bosons_scaling_high_T} in Fig.~\ref{periodic_scaling}(b). Thus, we establish effective thermalization of the localized state at finite temperatures.

	\section{Quasiperiodic case}
	\subsection{MBDL transition at finite temperatures}
	%%%%%%%%%%%%%%%%%%%%%%%%%%%%%%%%%%%%%%%%%%%%%%%%%%%%%%%%%%%%%
	\begin{figure}
		\centering
		\includegraphics[width=0.7\linewidth]{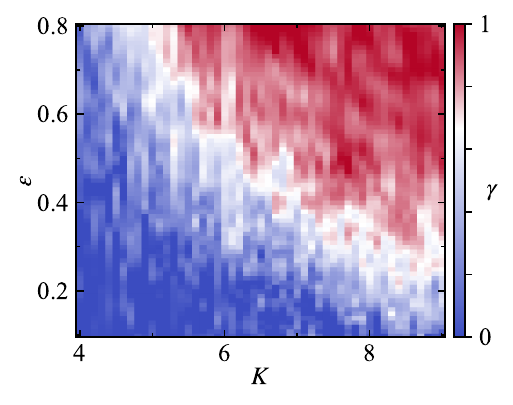}
		\caption{
			The extracted dynamical exponent $\gamma$ as a function of the anisotropy $\varepsilon$ and the kick strength $K$. The blue (red) regime denotes the localized (delocalized) phase, while the white regime denotes the anomalous diffusion with $\gamma\approx2/3$. The data are computed from the single-particle quasiperiodic QKR at $T_0=0.55\varepsilon_{\mathrm{F}}$. The particle number, and the effective Planck constant are respectively $N=1, \hbar_{\mathrm{eff}}=2.89$.
		}
		\label{quasiperiodic_phase}
	\end{figure}
	%%%%%%%%%%%%%%%%%%%%%%%%%%%%%%%%%%%%%%%%%%%%%%%%%%%%%%%%%%%%%
	
	%%%%%%%%%%%%%%%%%%%%%%%%%%%%%%%%%%%%%%%%%%%%%%%%%%%%%%%%%%%%%
	\begin{figure*}
		\centering
		\includegraphics[width=0.7\linewidth]{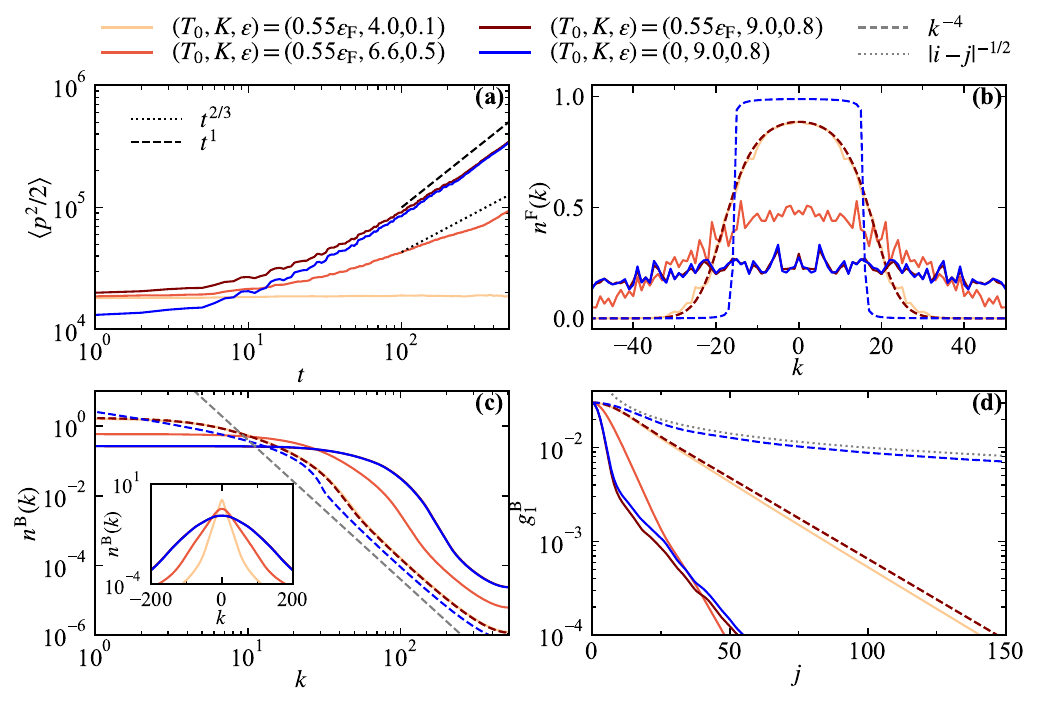}
		\caption{
			(a) The kinetic energy dynamics of the quasiperiodically kicked Tonks gas for different $(K,\varepsilon)$. (b-d) The corresponding fermionic momentum distribution (b), the bosonic momentum distribution (c), and the bosonic correlation function (d) for different $(K,\varepsilon)$. The dashed lines (blue and red) denote the initial state at $t=0$ while the solid lines denote the final state at $t=500$. The particle number, and effective Planck constant are respectively $N=31, \hbar_{\mathrm{eff}}=2.89$.
		}
		\label{quasiperiodic_dynamics}
	\end{figure*}
	%%%%%%%%%%%%%%%%%%%%%%%%%%%%%%%%%%%%%%%%%%%%%%%%%%%%%%%%%%%%%
	\begin{figure*}
		\centering
		\includegraphics[width=0.75\linewidth]{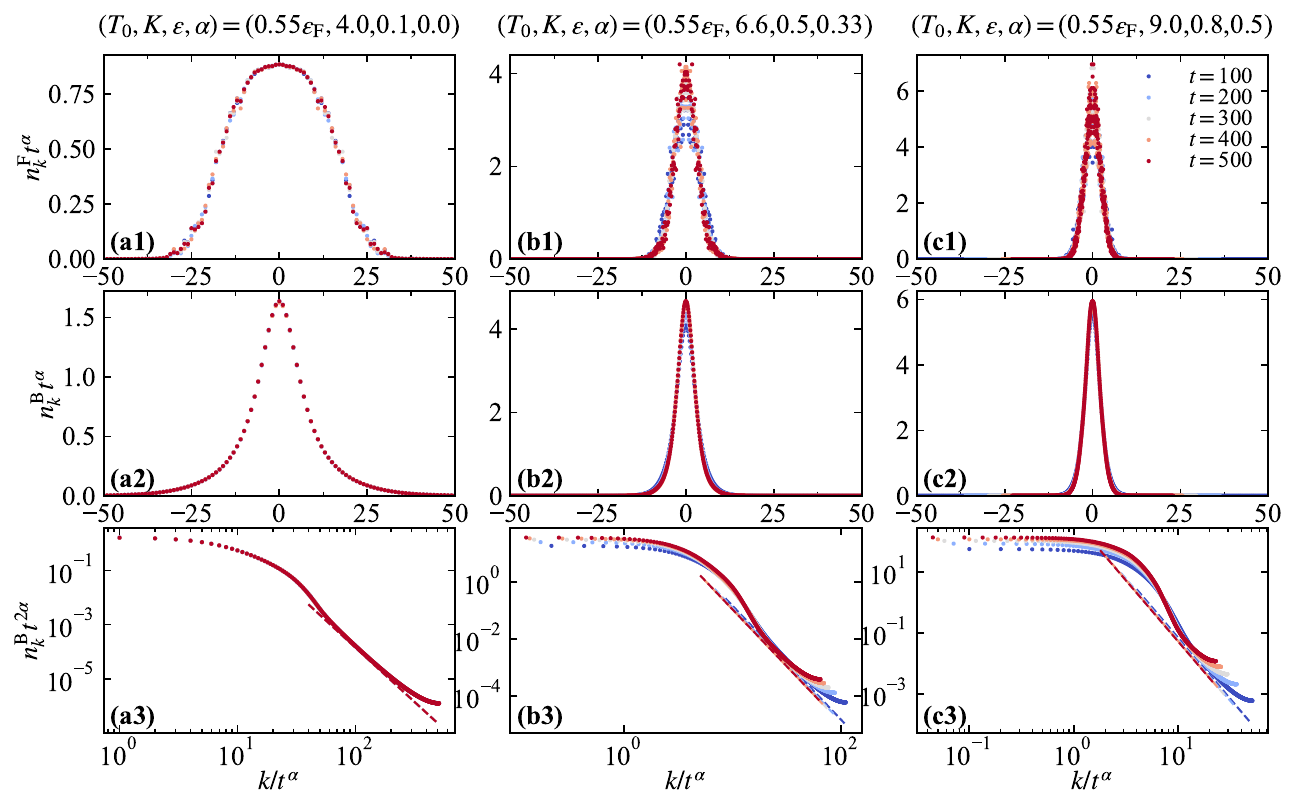}
		\caption{
			The rescaled fermionic and bosonic momentum distribution at different time $t$ in the localized (a1-a3), critical (b1-b3) and delocalized (c1-c3) phases. The dashed lines in (a3), (b3) and (c3) denote the predicted algebraic decay $\propto\mathcal{C}_{\mathrm{th}}k^{-4}$. The particle number, initial temperature, and effective Planck constant are respectively $N=31, T_0=0.55\varepsilon_{\mathrm{F}}, \hbar_{\mathrm{eff}}=2.89$.
		}
		\label{quasiperiodic_dynamics_scaling}
	\end{figure*}
	
	We have studied the many-body dynamical localization (MBDL) phase at finite temperatures in the above section. In the following, we focus on the delocalized phase at $\varepsilon\neq0$ and $T_0\neq0$, i.e., the quasiperiodically kicked Tonks-Girardeau (TG) gas. It was shown that the dynamics of the quasiperiodic Quantum kicked rotor (QKR) (Eq.~\eqref{total_hamiltonian} for $N=1$) is equivalent to that of a three-dimensional periodically kicked pseudo rotor starting from a special state~\cite{casatiquasi1989,delande2009metalinsulator},
	\begin{equation}
		\begin{aligned}
			\hat{H}_{\mathrm{sp}}(t)&=\frac{p_1^2}{2}+\omega_2p_2+\omega_3p_3 \\
			&+K\cos{x_1}[1+\varepsilon\cos{x_2}\cos{x_3}]\sum_n\delta(t-n),
		\end{aligned}
	\end{equation}
	where subscript ``sp" denotes single-particle. The above 3D pseudo rotor can be mapped onto a 3D Anderson-like model~\cite{casatiquasi1989}
	\begin{equation}
		\epsilon_{\mathbf{k}}\Phi_{\mathbf{k}}+\sum_{\mathbf{m}\neq0}W_{\mathbf{m}}\Phi_{\mathbf{k-m}}=-W_0\Phi_{\mathbf{k}},
		\label{3D_mapping}
	\end{equation}
	with the on-site term as
	\begin{equation}
		\epsilon_{\mathbf{k}}=\tan{\left\{\frac{1}{2}\bigg[\omega-\bigg(\hbar_{\mathrm{eff}}\frac{k_1^2}{2}+\omega_2k_2+\omega_3k_3\bigg)\bigg]\right\}}
	\end{equation}
	and the hopping term as
	\begin{equation}
		W_{\mathbf{k}}=\iiint \tan{\left[K\cos{x_1}(1+\varepsilon\cos{x_2}\cos{x_3})/2\hbar_{\mathrm{eff}}\right]}d\mathbf{x}^3,
	\end{equation}
	where $\mathbf{k}=(k_1,k_2,k_3)$ denotes the 3D momentum vector. When the set $(\hbar_{\mathrm{eff}}, \omega_2, \omega_3, \pi)$ is incommensurate, the potential $\epsilon_{\mathbf{k}}$ behaves pseudorandomly. Moreover, $W_{\mathbf{k}}$ is a short-range hopping, so Eq.~\eqref{3D_mapping} is an effective 3D Anderson model, and the predicted Anderson transition is expected. Note that the strength of pseudorandomness $\epsilon_{\mathbf{k}}$ is fixed; we expect that increasing $K$ and $\varepsilon$, thus increasing the hopping strength $W_{\mathbf{k}}$, will lead to delocalization.
	
	As we have recalled in the Model section, at zero temperature, the energy of the Tonks gas is the same as that of free fermions, which is the sum of each fermion. Thus, the Tonks gas shares the same phase diagram as for free fermions~\cite{vuatelet2021effectivethermal,Vuatelet2023dynamicalmanybody}. At finite temperature, the quasiperiodic QKR is not yet systematically explored. The Jordan-Wigner transformation still ensures the same Hamiltonian form of the Tonks gas and free fermions~\cite{rigol2005_thermalTG,rigol2017thermalexpansion}, and thus the same local observables. Since free fermions are noninteracting, we first thoroughly quantify different phases of the quasiperiodic QKR for $N=1$ at finite temperature, and in different regions of parameter space $(K,\varepsilon)$. Then we choose the parameters of interest to study the Tonks gas for $N>1$, and the same dynamical behavior should be expected. Although the energy values of many particles differ from those of the single-particle case, the dynamical growth behavior is the key factor determining localization/delocalization. 
 
     To do this, we introduce the dynamical exponent $\gamma$ defined as
	\begin{equation}
		\gamma=\lim_{t\rightarrow\infty}\frac{d\log{\langle p^2/2\rangle}}{d\log{t}}.
	\end{equation}
	The kinetic energy dynamics is considered as diffusion for $\gamma=1$, anomalous diffusion for $\gamma=2/3$, and localized for $\gamma\sim0$. In Fig.~\ref{quasiperiodic_phase}, for the single-particle case $N=1$, we scan the dynamical exponents $\gamma$ for different $K$ and $\varepsilon$ at an intermediate initial temperature $T_0=0.55\varepsilon_{\mathrm{F}}$. As expected, there are three phases, i.e., localized (blue), critical (white), and delocalized (red). Thus, we choose three typical parameters, $(K, \varepsilon)=(4.0,0.1)$, $(K, \varepsilon)=(6.6,0.5)$, and $(K, \varepsilon)=(9.0,0.8)$ to study the kicked Tonks gas in the three phases above, respectively.
	
	Next, we present the many-body results. As shown in Fig.~\ref{quasiperiodic_dynamics}(a), the kinetic energy of the thermal Tonks gas grows diffusively ($p^2/2\sim t$) in the delocalized phase, follows an anomalous diffusion ($p^2/2\sim t^{2/3}$) in the critical phase, and hardly grows in the MBDL phase. The dynamical exponents align with what is expected at these parameters in the single-particle phase diagram Fig.~\ref{quasiperiodic_phase}. In the delocalized phase, the system continuously absorbs energy, eventually approaching an infinite-temperature state; consequently, the initial temperature becomes negligible at late times. This can be seen from the fact that, in the delocalized phase, except at early times, the dynamics at zero temperature (blue solid line) closely follow that at $T_0=0.55\varepsilon_{\mathrm{F}}$, and the other observables behave similarly. For long enough evolution, this indicates that finite temperature does not affect the MBDL transition. From the fermionic momentum distributions, one can see that $n^{\mathrm{F}}(k)$ is extremely broad in the delocalized phase. This is much clearer from the bosonic momentum distribution, where $n^{\mathrm{B}}(k)$ at low momentum changes from an exponential decay to a broader one and finally to a Gaussian form, as the parameters change (inset in Fig.~\ref{quasiperiodic_dynamics}(c)). This transition from localization to diffusion is a process in which the coherence is gradually destroyed. As shown in Fig.~\ref{quasiperiodic_dynamics}(d), the correlation function $g^{\mathrm{B}}_1(r)$ decays more rapidly as the system crosses from localized to delocalized phases.
	
	\subsection{The scaling laws in different phases}
	The 1D quasiperiodic Quantum kicked rotor (QKR) belongs to the orthogonal universality class of the 3D Anderson model~\cite{delande_2009universality,slevin1999andersonscaling,rodriguez2010criticalanderson}, so the momentum distribution in each phase obeys a distinct one-parameter scaling law~\cite{Ramakrishnan1979scalingtheory}. Here, we mainly focus on the moderate and large momentum region. From the energy dynamics, we can infer that the momentum grows with time in the form of $p\sim t^{1/2}$ in the delocalized phase but $p\sim t^{1/3}$ in the critical phase. Therefore, we expect the following scaling function
	\begin{equation}
		n(k)=t^{-\alpha}f(kt^{-\alpha}),
		\label{moderate_k_scaling}
	\end{equation}
	with 
	\begin{equation}
		\alpha= \frac{\gamma}{2} = \left\{
		\begin{aligned}
			& 1/2 & & \mathrm{delocalized}, \\
			& 1/3 & & \mathrm{critical}, \\
			& 0 & & \mathrm{localized}.
		\end{aligned}
		\right.
	\end{equation}
	The rescaled momentum distributions at moderate momenta in three phases are shown in Fig.~\ref{quasiperiodic_dynamics_scaling}(a1-a2), (b1-b2), and (c1-c2). At late times, both the fermionic and bosonic momentum distributions collapse very well to the corresponding scaling function. However, at very large momenta, the bosonic momentum distribution holds an algebraic decay $\propto\mathcal{C}_{\mathrm{th}}k^{-4}$, which does not obey the scaling Eq.~\eqref{moderate_k_scaling}. Similar to the zero temperature case, there is another scaling function for the algebraic tail. Notice that $\mathcal{C}_{\mathrm{th}}\propto p^2$ and thus $\mathcal{C}_{\mathrm{th}}\propto t^{2\alpha}$. Therefore, we expect the following scaling function
	\begin{equation}
		n(k)=t^{-2\alpha}F(kt^{-\alpha}).
		\label{large_k_scaling}
	\end{equation}
	As shown in Fig.~\ref{quasiperiodic_dynamics_scaling}(a3), (b3), and (c3), the rescaled algebraic tails as well as the predicted decay (dashed lines) are well captured by Eq.~\eqref{large_k_scaling}.
	
	\subsection{Effective thermalization of the delocalized state at finite temperatures}
    %%%%%%%%%%%%%%%%%%%%%%%%%%%%%%%%%%%%%%%%%%%%%%%%%%%%%%%%%%%%%
	\begin{figure}
		\centering
		\includegraphics[width=0.85\linewidth]{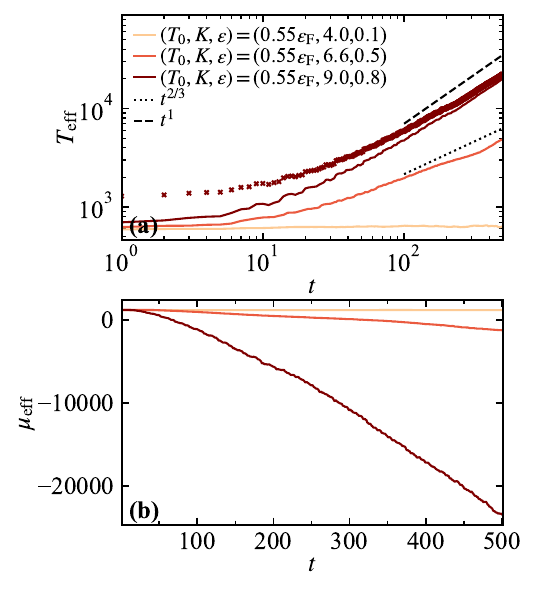}
		\caption{
			The dynamics of the extracted effective temperature $T_{\mathrm{eff}}$ (a) and the chemical potential $\mu_{\mathrm{eff}}$ (b) for different $(K,\varepsilon)$. The crossings denote the prediction of the equipartition theorem. The particle number, initial temperature, and effective Planck constant are respectively $N=31, T_0=0.55\varepsilon_{\mathrm{F}}, \hbar_{\mathrm{eff}}=2.89$.
		}
		\label{quasiperiodic_fitting}
	\end{figure}
    %%%%%%%%%%%%%%%%%%%%%%%%%%%%%%%%%%%%%%%%%%%%%%%%%%%%%%%%%%%%%
	\begin{figure}
		\centering
		\includegraphics[width=0.75\linewidth]{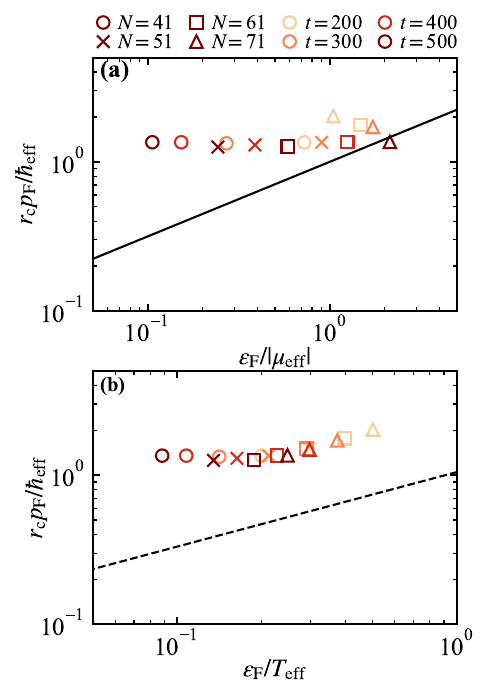}
		\caption{
			$r_{\mathrm{c}}p_{\mathrm{F}}/\hbar_{\mathrm{eff}}$ as a function of $\varepsilon_{\mathrm{F}}/\left|\mu_{\mathrm{eff}}\right|$ (a) and $\varepsilon_{\mathrm{F}}/\left|T_{\mathrm{eff}}\right|$ for different particle numbers $N$ and different times $t$. The black solid and dashed lines show the predictions for $\mu_{\mathrm{eff}}/T_{\mathrm{eff}}\gg 1$ and $\mu_{\mathrm{eff}}/T_{\mathrm{eff}}\ll 1$, respectively. The anisotropy, kick strength and effective Planck constant are respectively $\varepsilon=0.8, K=9, \hbar_{\mathrm{eff}}=2.89$.
		}
		\label{quasiperiodic_scaling}
	\end{figure}
	%%%%%%%%%%%%%%%%%%%%%%%%%%%%%%%%%%%%%%%%%%%%%%%%%%%%%%%%%%%%%
	In the last section, we have studied the effective thermalization at different $T_0$ in the many-body dynamical localization (MBDL) phase. Yet in the delocalized phase, the effect of $T_0$ is negligible since the system is at infinite temperatures. Same as before, we extract the effective temperature $T_{\mathrm{eff}}$ and the effective chemical potential $\mu_{\mathrm{eff}}$ by imposing the constraints Eq.~\eqref{constraints} to the fermionic momentum distribution $n^{\mathrm{F}}(k)$, as shown in Fig.~\ref{quasiperiodic_fitting}. Notably, even within the delocalized and critical phases, $T_{\mathrm{eff}}$ scales with energy growth according to the power law $T_{\mathrm{eff}}\sim t^{2\alpha}$. We notice that in the delocalized phase, $\mu_{\mathrm{eff}}$ is negative at late times with high $T_{\mathrm{eff}}$. Therefore, the Fermi-Dirac distribution changes to the Maxwell-Boltzmann distribution, and the fermions are non-degenerate. We can use the equipartition theorem to build a relation between $T_{\mathrm{eff}}$ and the kinetic energy as
	\begin{equation}
		\left\langle\frac{p^2}{2}\right\rangle=\frac{NT_{\mathrm{eff}}}{2}, 
		\label{equipartition}
	\end{equation}
	thus we obtain $T_{\mathrm{eff}}=\langle p^2\rangle/N$. The crossings in Fig.~\ref{quasiperiodic_fitting}(a) are the prediction of Eq.~\eqref{equipartition}. At late times, deep in the delocalized regime, $T_{\mathrm{eff}}$ is accurately described by this prediction. For a thermal Tonks gas with negative chemical potential ($\mu_{\mathrm{eff}}<0$), the correlation length of the correlation function $g^{\mathrm{B}}_1$ is known to read~\cite{korepin1991correlation}
	\begin{equation}
		r_{\mathrm{c}}^{-1}=\frac{\sqrt{2|\mu_{\mathrm{eff}}|}}{\hbar_{\mathrm{eff}}}+\frac{\sqrt{2k_{\mathrm{B}}T}}{2\pi\hbar_{\mathrm{eff}}}\int_{-\infty}^{\infty}d\lambda \left|\frac{e^{\lambda^2-\mu_{\mathrm{eff}}/T}+1}{e^{\lambda^2-\mu_{\mathrm{eff}}/T}-1}\right|.
	\end{equation}
	The difference of $r_{\mathrm{c}}^{-1}$ between $\mu_{\mathrm{eff}}>0$ and $\mu_{\mathrm{eff}}<0$ cases is just the effective chemical potential. For $|\mu_{\mathrm{eff}}|/T_{\mathrm{eff}}\gg1$, we have the following scaling~\cite{korepin1991correlation,Vuatelet2023dynamicalmanybody}
	\begin{equation}
		\frac{r_{\mathrm{c}}p_{\mathrm{F}}}{\hbar_{\mathrm{eff}}}=\sqrt{\frac{\varepsilon_{\mathrm{F}}}{|\mu_{\mathrm{eff}}|}},
		\label{delocalized_lowT}
	\end{equation}
	but for $|\mu_{\mathrm{eff}}|/T_{\mathrm{eff}}\ll1$, we return to Eq.~\eqref{bosons_scaling_high_T}. Note that, in the evolution time we consider ($t\leq500$), $T_{\mathrm{eff}}$ is slightly larger than $|\mu_{\mathrm{eff}}|$. Consequently, in the delocalized phase, the data follow Eq.~\eqref{bosons_scaling_high_T} while deviating from Eq.~\eqref{delocalized_lowT}, as shown in Fig.~\ref{quasiperiodic_scaling}.

	\section{Discussions}
	In summary, we have investigated the dynamics of a thermal Tonks-Girardeau (TG) gas subjected to periodic and quasiperiodic kicks. Our results show the persistence of the many-body dynamical localization (MBDL) at finite and even high temperatures. Finite temperatures increase the system's kinetic energy and further degrade the coherence of the TG gas. We establish effective thermalization of the MBDL state, with modifications arising from the nonzero initial temperature. Finally, we show the MBDL transition at an intermediate temperature and different scaling laws in different phases. Except for the early times, the delocalized phase at late times is not affected by the initial temperature. Our study complements the properties of the MBDL and delocalized phases at finite temperatures and is expected to provide guidance and explanation for experiments observing these phases at finite temperatures.

    The Lieb-Liniger parameter $\gamma=gL/(N\hbar^2_{\mathrm{eff}})$ is generally recognized as an intensive parameter representing the dimensionless interaction~\cite{lieb1963solution1}. Different interaction regimes are characterized by their value, e.g., $\gamma\ll1$ (mean-field regime), and $\gamma\gg1$ (strongly-interacting or TG regime). In the latter case, interaction is strong enough to cause on-site repulsion in real space, thus making hard-core bosons a good description. Our study focuses on the solvable TG limit $g\rightarrow\infty$, which has been experimentally verified to perfectly describe the dynamic properties of a Bose gas at $\gamma\gg1$~\cite{paredes2004tonks,wilson2020observation,guo2023observation}. Further considering the effect of initial temperature, our results enable accurate predictions and serve as a benchmark for low-temperature cold-atom experiments in the strongly-interacting regime.
	
	It is worth noting that our study is limited to the case of infinite interactions, though in which many particles are exactly considered, local properties of the system are determined by the underlying free fermions. A wider variety of phenomena should emerge in a genuinely finite-interaction many-body setting. Finite interactions, particularly at intermediate strengths, introduce long-range correlations that may induce thermalization~\cite{olsen2025anderson,yang2025origin}. While methods such as the Bethe-Ansatz or second-quantization treatments of the Hamiltonian in Eq.~\eqref{total_hamiltonian} can handle the cases with finite interactions, applying these to many-particle systems at finite temperatures remains numerically challenging due to the non-integrability. Therefore, further theoretical and numerical research is essential to explore many-body dynamical localization in the finite-interaction regime.
	
	\section*{Acknowledgments}
	This work was supported by the National Natural Science Foundation of China (No. 12375021), the Zhejiang Provincial Natural Science Foundation of China (No. LD25A050002), the National Key Research and Development Program of China (No. 2022YFA1404203).

\end{document}